\documentclass[aps,prc,preprint,superscriptaddress,a4paper,showkeys]{revtex4}

\usepackage{amsmath}
\usepackage{amssymb}
\usepackage[dvipsnames,usenames]{color}
\usepackage[pdftex]{graphicx,epsfig}

\usepackage{lmodern}
\usepackage[T1]{fontenc}
\usepackage[english]{babel}

\usepackage[caption=false]{subfig}
\usepackage{epstopdf}
\usepackage{bm}
\usepackage{wasysym}
\usepackage{textcomp}
\usepackage{gensymb}
\usepackage{ulem}
\usepackage{xspace}
\usepackage[tight,nice]{units}
\usepackage[breaklinks=true,hidelinks]{hyperref}

\newcommand{\ts}[1]{#1}
\newcommand{\tsdel}[1]{}

\newlength{\bildtitel}
\newcommand{\mrm}{\mathrm}

\newcommand{\diff}[1]{\operatorname{d}\ifthenelse{\equal{#1}{}}{\,}{\!#1}}

\newcommand{\ueV}{\mbox{$\micro$eV}}

\begin{document}

\title{Experimental study of ultracold neutron production in pressurized superfluid helium}

\author{P.~Schmidt-Wellenburg}
\email[Corresponding author:~]{philipp.schmidt-wellenburg@psi.ch}
\altaffiliation[\newline $^{\ast}$New
affiliation: ]{Paul Scherrer Institut, 5232 Villigen PSI, Switzerland}
\affiliation{Institut Laue Langevin, BP 156, 38042
Grenoble, France}\affiliation{Physik-Department E18, Technische
Universit\"at M\"unchen, D-85748 Garching, Germany}

\author{J.~Bossy}\affiliation{Institut N\'eel, CNRS-UJF, BP 38042 Grenoble Cedex 9, France}
\author{E.~Farhi}\affiliation{Institut Laue Langevin, BP 156, 38042
Grenoble, France}
\author{M.~Fertl}\altaffiliation[New affiliation: ]{University of Washington, Seattle, United States of America}\affiliation{Physik-Department E18,
Technische Universit\"at M\"unchen, D-85748 Garching, Germany}
\author{K.K.H.~Leung}\altaffiliation[New affiliation: ]{Department of Physics,
  North Carolina State University, Raleigh, USA}\affiliation{Institut Laue Langevin,
  BP 156, 38042 Grenoble, France}\affiliation{Physik-Department E18, Technische
  Universit\"at M\"unchen, D-85748 Garching, Germany}
\author{A.~Rahli}\affiliation{Universit\'e Mouloud Mammeri, 15000 Tizi-Ouzou, Algerie}
\author{T.~Soldner}
\email[Corresponding author:~]{soldner@ill.eu}
\affiliation{Institut Laue Langevin, BP 156, 38042 Grenoble, France}
\author{O.~Zimmer}
\affiliation{Institut Laue Langevin, BP 156, 38042 Grenoble, France}
\affiliation{Physik-Department E18, Technische Universit\"at
M\"unchen, D-85748 Garching, Germany}
\date{\today}

\begin{abstract}
    We have investigated experimentally the pressure dependence of the production of
    ultracold neutrons~(UCN) in superfluid helium in the range
    from saturated vapor pressure to \unit[20]{bar}.
    A neutron velocity selector allowed the separation of
    underlying single-phonon and multiphonon processes by
    varying the incident cold neutron~(CN) wavelength in the range from $3.5$ to \unit[10]{\AA}.
    The predicted pressure dependence of UCN production derived from
    inelastic neutron scattering data was confirmed for the
    single-phonon excitation. For multiphonon based UCN
    production we found no significant dependence on pressure whereas
    calculations from inelastic neutron scattering data predict an
    increase of \unit[43(6)]{\%} at \unit[20]{bar} relative to saturated vapor pressure.
    From our data we conclude that applying pressure to superfluid
    helium does not increase the overall UCN production rate at a
    typical CN guide.
\end{abstract}

\maketitle
\section{Introduction}
Ultracold neutrons (UCN) have energies below \unit[300]{neV} and can be stored for 
long observation time in magnetic or material bottles\,\cite{Golub1991}. This peculiarity makes them
attractive for high precision measurements of
fundamental properties of the neutron which are
relevant for particle physics and cosmology\,\cite{Dubbers2011}.
The most prominent example is their use for searches of a neutron electric dipole
moment\,\cite{Baker2006,Pospelov2005,Lamoreaux2009,Baker2011,Altarev2012,Serebrov2014} which would give direct evidence of \textit{CP}-violation beyond the Standard Model of
particle physics.
Other projects aim to improve the knowledge of the neutron $\beta$-decay lifetime which is crucial for calculations of big bang
nucleosynthesis\,\cite{Coc2007}. Together
with correlation measurements in neutron decay it contributes to a
more precise
understanding of weak semileptonic processes in the first quark
generation\,\cite{Severijns2006,Abele2008}. New developments might
follow from observations and manipulations of quantum states of UCN in
the Earth's gravitational field\,\cite{Nesvizhevsky2002,Jenke2011,Jenke2014}.
All of these endeavors greatly benefit from an increase of UCN counting statistics in their experiments.

Current projects to increase the density of UCN for physics
experiments employ neutron converters of superfluid helium
(He-II)\,\cite{Huffman2000,Masuda2002,Baker2003,Zimmer2007,Zimmer2010,Piegsa2014}
and solid deuterium\,\cite{Trinks2000,Anghel2009,Korobkina2007,Saunders2013,Lauer2013,Lauss2013}.
They exploit the concept of a superthermal source \cite{Golub1975}
where the converter temperature can be much higher than the UCN ``temperature''.
Upscattering of UCN is suppressed by the Boltzmann factor if the energy of
excitations in the converter is large compared to its temperature. Using liquid He as
converter was proposed in \cite{Golub1977} and first experimentally realized in
\cite{Ageron1978}.

In He-II the main contribution to UCN production is due
to single-phonon excitations (see \cite{Baker2003} for a separation
of UCN production in single- and multiphonon contributions), which occur
at the crossing point $\lambda^\ast$ of
the dispersion relations of the free neutron
and of He-II\@.
At saturated vapor pressure~(SVP), $\lambda^\ast = \unit[8.92(2)]{\text{\AA{}}}$
(derived from the data in~\cite{Gibbs1999},\footnote{Note that a more
accurate value for SVP may be derived from the comprehensive data collection
\cite{Donnelly1998}. However, we wanted to base our comparisons with
scattering data on one consistent data set.}).
Application of pressure shifts $\lambda^\ast$ to lower values, where the differential flux $\mathrm{d}\phi
/\mathrm{d}\lambda$ is higher for a typical neutron beam in a guide
coupled to a liquid deuterium CN source, which typically has its flux maximum around \unit[4]{\AA}.
Combined with an increase in density of He-II with pressure one may anticipate an overall gain
in UCN production.

However, a quantitative analysis based on a method described in
Ref.\,\cite{Korobkina2002} using inelastic neutron scattering data
at SVP and \unit[20]{bar} from \cite{Gibbs1999} predicted a net
decrease in single-phonon UCN
production\,\cite{Schmidt-Wellenburg2009,Schmidt-WellenburgPhD}.
On the other hand, these same calculations also predicted an increase in
the production from multiphonon excitations at \unit[20]{bar}. We wished
to test these calculations and quantify the UCN
production in pressurized He-II by a direct measurement.

Independent of UCN production, another motivation for performing these studies is the larger dielectric strength of pressurized He-II\,\cite{Hara1998}.
This is of particular interest for searches of the neutron electric dipole
moment within He-II using electrical fields of several
\unit[10]{kV/cm}\,\cite{Harris2010}.

\section{Experimental setup}
\begin{figure}
    \centering
    \includegraphics[width=1\columnwidth]{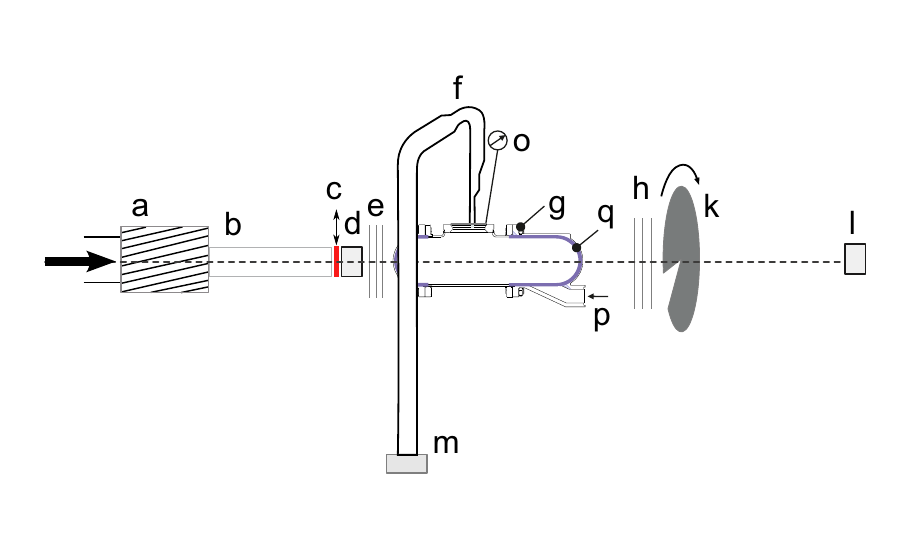}

    \caption{(Color online) Schematic of the setup (not to scale): The CN~beam passed through
      the velocity selector~a), a secondary CN guide~b) which could be closed
      off with a fast shutter~c),
      and a beam monitor with circular orifices~d) before it entered the cryostat
      through a series of thin aluminum windows~e). The He-II target was
      kept in the Ni coated stainless steel pressure container~g); CN entrance and exit sides from aluminum were covered with nickel half-spheres~q). Ultracold neutrons were guided by a polished stainless steel tube~f),
      which was separated by an aluminum window
      from the production volume, to the $ ^3$He UCN detector~m). The
      CN beam exited the cryostat through aluminum windows~h) before
      entering a chopper~k) and TOF-detector~l) for time-of-flight spectral
      analysis. Pressure was applied via the filling line~p) connected on the bottom of the pressure container and measured on a static line with a gauge~o).}
    \label{fig:setup}
\end{figure}

The experiment took place at the CN~beam
facility PF1b\,\cite{Abele2006} of the Institut Laue-Langevin, France. A schematic diagram of the experimental setup is
shown in Fig.\,\ref{fig:setup}. The setup for the He-II UCN
production target and UCN extraction was similar to the one
described in \cite{Zimmer2007,Zimmer2010,Piegsa2014}: The target was located
inside a cryostat equipped with a commercial two-stage Gifford McMahon
cold head with a cooling power of \unit[1.5]{W} at \unit[4.2]{K}. In the cryostat,
gaseous He with 99.999\% purity was liquefied and cooled below the
$\lambda$-transition to superfluidity by a continuous $^4$He
evaporation stage. $^3$He was removed by passing the superfluid through a
superleak (for technical details on the used design, see \cite{Zimmer2010})
into the UCN production volume. Ultracold neutrons were extracted by an
inverted U-shaped stainless steel UCN guide to a $^3$He UCN
detector. For further details, see\,\cite{Zimmer2007,Piegsa2014,Schmidt-WellenburgPhD}.

In order to allow for measurements at high pressure, a dedicated He-II container was used as converter vessel. It consisted of a stainless steel tube (inner diameter \unit[66]{mm}, length \unit[193.8]{mm}) with two spherical end caps from aluminum (radius $\sim\unit[33]{mm}$, wall thickness \unit[3]{mm}) for passage of the CN beam. Its total volume was $\unit[810]{cm^3}$ and the intersection volume with the neutron beam about $\unit[180]{cm^3}$.
The cylinder was coated with natural nickel and the end caps had nickel half-spherical shells inserted on the inside to increase the neutron
optical potential for UCN storage. The UCN flapper valve used
in\,\cite{Zimmer2007} for window-less extraction was replaced by an aluminum window ($\unit[0.1]{mm}$ thickness) supported by a
stainless steel disk with an array of holes (effective hole area
$127~\mrm{mm^3}$).  
The $^3$He evaporation
stage serving to further cool the production volume had to be
operated with $^4$He due to the higher heat load of the high
pressure volume. Therefore the target temperature reached only \unit[1.1]{K}
instead of \unit[0.7]{K} obtained in \cite{Zimmer2007}. The temperature
of the He-II was measured using a calibrated
Cernox\footnote{Model: CX-1030-CU, Lake Shore Cryogenic, Inc.}
sensor screwed tightly to the outside of the He-II 
vessel. 
The pressure was applied from the gaseous He supply via
the liquefaction chain through the superleak and measured with a room-temperature pressure gauge calibrated for $p<\unit[100]{bar}$ relative to atmospheric pressure. The gauge was connected to the converter via a thin stainless steel tube which could also 
be used for pumping to establish SVP conditions.

\begin{figure}
    \centering
    \includegraphics[width=1\columnwidth]{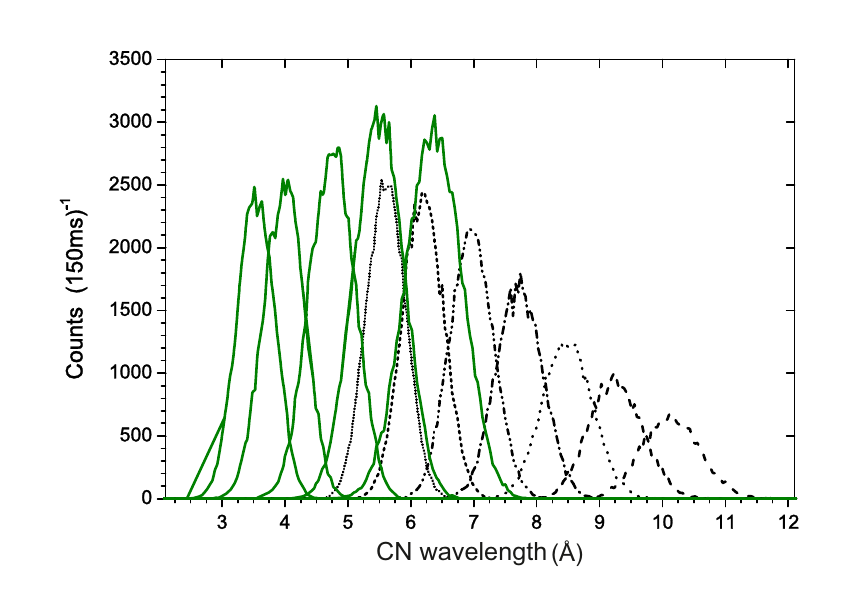}
    \caption{(Color online) Time-of-flight spectra of the CN beam measured with empty
      production volume, for various velocity selector rotation speeds
      at $-5^{\circ}$ ( {\color{OliveGreen}\textbf{---}} ) and
      $+5^{\circ}$ ( -\,-\,- ) selector alignment.
     Each raw spectrum was normalized to area unity and then multiplied with the respective monitor count rate corrected for dead time and $1/v$ efficiency. The difference
     in integral counts for $-5^{\circ}$ and $+5^{\circ}$ was due to
     a larger transmission of the velocity selector at $-5^{\circ}$.}
    \label{fig:spectrum}
\end{figure}

A neutron velocity
selector\,\cite{Wagner1992} was used to scan the incident CN~wavelength range. 
It had to be used in two orientations,
$-5^{\circ}$ and $+5^{\circ}$ with respect to the incident beam, as
the normal position ($0^{\circ}$) did not give access to the entire
wavelength range of interest. Hence, two scans were performed: from
$3.5$~\AA{} to $6.5$~\AA{} and from $5.5$~\AA{} to $10$~\AA{}.
Depending on the angular and rotational settings of
the velocity selector a wavelength spread of 
$\tfrac{\Delta\lambda\left(\mrm{FWHM}\right)}{\lambda} \approx 0.14,~0.08$
was obtained for $-5^{\circ}$, $+5^{\circ}$ respectively. After passage of
a $2.5$~m long secondary neutron guide (cross section $30 \times 50$~mm$^2$)
with natural nickel coating, a fast shutter in front of the beam monitor
was used to open and close the beam (opening and closing time less than \unit[1]{s}).
Before entering the beam monitor the beam's cross section
was reduced to $\varnothing = \unit[30]{mm}$ with two circular apertures
made from B$_4$C. Images of the beam at various
distances behind the monitor were taken before the cryostat was set
up. They showed that the beam passed through the production volume
without touching its cylindrical sidewall for all selected incident
wavelengths. A time-of-flight (TOF) system for
spectral analysis of the incident spectrum was set up behind the
cryostat. It consisted of a chopper
--- a disc rotating at \unit[1000]{RPM} with four radial rectangular slits (width \unit[3]{mm}) passing by
a fixed aperture of the same dimensions ---  and a $ ^3$He CN-monitor with a
detection efficiency proportional to $1/v$ ($v$ is the neutron
velocity) placed \unit[2]{m} downstream. The corresponding wavelength
spectra for some selected velocity selector settings are displayed
in Fig.\,\ref{fig:spectrum}.

\section{Measurements}
\begin{figure}
    \centering
    \subfloat{
    \includegraphics[width=0.45\columnwidth]{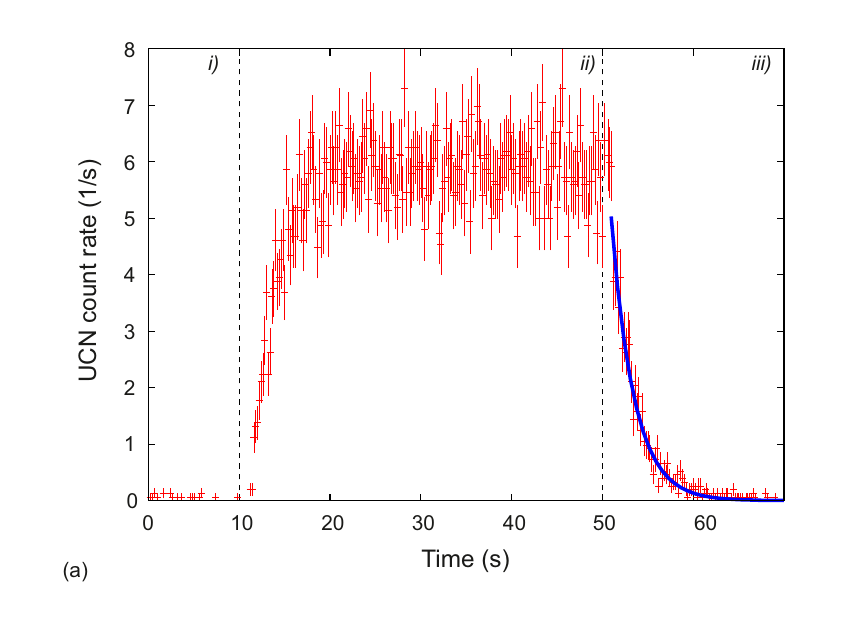}}
		\subfloat{
		\includegraphics[width=0.45\columnwidth]{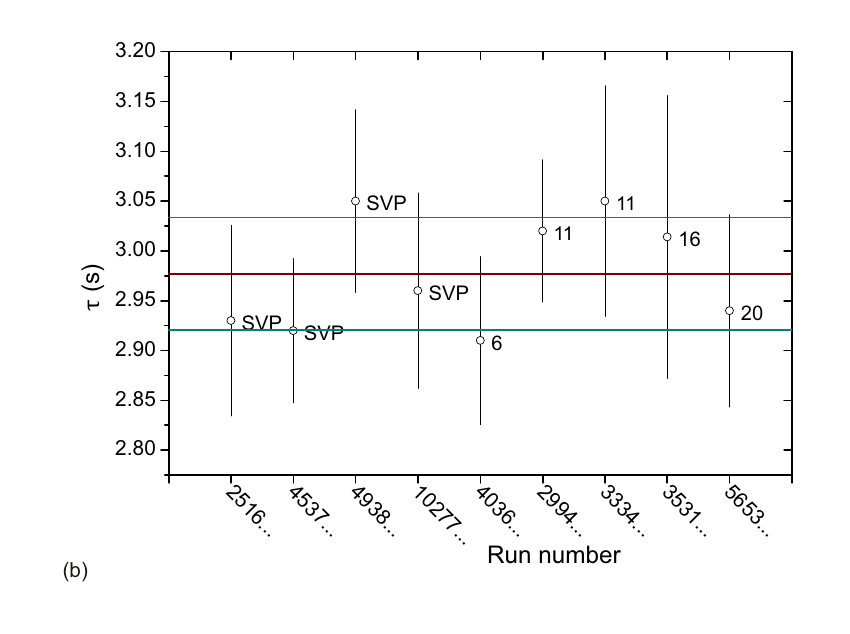}}
    \caption{(Color online) 
		Count rate observed in the UCN detector~(a). Shown is the combined data of all
    measurements at SVP and $\lambda \sim \unit[8.7]{\text{\AA}}$
    (maximum of UCN production, $\lambda_\text{cor} = \unit[8.92]{\text{\AA}}$). 
		The vertical
    lines indicate the three different phases. 
		The solid line in phase iii) is the best fit (see text)
    to the data with a reduced $\chi^2=1.04$.
		The decay time constants $\tau$ for different pressure were constant 
		within the error bars~(b): The red line~({\bf \color{Red} ---}) is the weighted mean, the blue lines~({\bf \color{BlueGreen} ---}) indicate the error of the weighted mean.
		}
    \label{fig:measurement}
\end{figure}

Scans of UCN count rate versus CN wavelength were performed at absolute pressures and temperatures
$(p\,[\mrm{bar}],T\,[\mrm{K}])= (6,1.105)$, $(11,1.106)$,
$(16,1.112)$, $(20,1.109)$, and (SVP, 1.08), with temperature
fluctuations $\Delta T \leq \unit[0.02]{K}$ during individual scans. The UCN
production rate was first measured in the long wavelength range with
the velocity selector angle set to $+5^{\circ}$ for all pressures.
Then the measurement of the UCN production rate in the short
wavelength range ($-5^{\circ}$) followed. For each pressure and
angular setting, three up -- down scans were performed with the same set of
velocity selector frequencies corresponding to wavelength increments
of about $0.25$~\AA{}\tsdel{ were performed}. During each scan the UCN
production rate for a specific rotation frequency was measured five
times. Each of these measurements was divided into three phases as
shown in Fig.\,\ref{fig:measurement}: 
i)~\unit[10]{s} with CN-shutter closed to verify the absence of UCN in the converter vessel; 
ii)~\unit[40]{s} with CN-shutter open to record the build-up and saturation of the UCN rate;
iii)~\unit[20]{s} with CN-shutter closed to record the emptying of the converter vessel.
The decrease in rate during phase (iii) was fitted with an exponential function (see
Fig.\,\ref{fig:measurement}):
$\dot{n}(t)=\dot{n}_0\exp\left(\tfrac{-t}{\tau}\right)+r_0$, where the fit parameter $\tau$
is the storage lifetime of the converter ($\tau=\unit[2.92(7)]{s}$ for SVP at
$T=\unit[1.08(2)]{K}$). The beam-independent
background $r_0 = 0^{+3.9}_{-0.0}\times 10^{-3}~\mrm{s^{-1}}$ was determined in
dedicated measurements with CN-shutter closed. This parameter was
kept fixed in the fits.
It was found that the storage lifetime was independent of pressure (see Fig.\,\ref{fig:measurement}b).
Increasing the temperature to \unit[1.48]{K} decreased $\tau$ to
\unit[1.7(1)]{s}. This temperature behavior combined with results from UCN
Monte Carlo simulations with Geant4UCN~\cite{Atchison2005} and
simple gas kinetic arguments showed that the storage lifetime was
dominated by UCN escaping through the extraction holes. Hence,
changes in count rate due to the small temperature fluctuations of $\pm
\unit[0.02]{K}$ were negligible.

In phase (ii), the background was increased by
a beam-dependent component $r_{\text{beam}}$ causing a 
constant offset to the UCN build-up prompt with opening the CN
shutter. A systematic error \tsdel{would}\ts{will} follow if this offset is not
properly taken into account. The component $r_{\text{beam}}$ was
determined by fitting the build-up with an
exponential with constant background $r=r_0+r_{\text{beam}}$.
It was expected to be proportional to the CN monitor count rate since CN monitor and
UCN detector both had a low efficiency proportional to $1/v$ for
CN\@.
Therefore $r_{\text{beam}}$ was determined for four
different velocity selector settings (maximum of single-phonon
production and three settings with low ratio of UCN production to CN
monitor count rate). A fit of the four values with
$r_{\text{beam}}(\dot{n}_{\text{CN}}) = k \dot{n}_{\text{CN}},$ where
$\dot{n}_{\text{CN}}$ was the CN count rate of the beam
monitor ($12000~\mrm{s^{-1}}<\dot{n}_{\text{CN}}<17000~\mrm{s^{-1}}$), yielded
$k=(5.6^{+6.0}_{-5.6})\times10^{-7}$. All data of phase (ii) was corrected for
both background contributions, by typically $\unit[0.01]{s^{-1}}$.

In addition to these scans the UCN production rates across the
single-phonon peak at SVP \tsdel{was}\ts{were} measured several times during the
entire experiment to check the reproducibility.
The UCN count rate was found to decrease with time. This is suspected to be due to gas
freezing out on the thin aluminum window at the extraction holes or on the
surface of the extraction guide. An exponential decrease with
respect to the start of the first measurement ($t=0$):
$N(t)=N_0\exp\left(\tfrac{-t}{{\cal T}}\right)+A_0$ with a large
offset $A_0 =1.95(10)\times N_0$ and time constant ${\cal T} =
104(12)$~h was found to describe this effect well and thus was used
to correct the data. The corrected UCN counts 
were normalized to the particle flux as measured by the beam
monitor (monitor count rate is corrected for dead time and $1/v$
efficiency). Time-of-flight measurements (see
Fig.\,\ref{fig:spectrum}) were used to determine a central wavelength
for each setting of the velocity selector.
The measured TOF spectra were deconvoluted taking into
account the opening function of the chopper and the finite thickness
of the TOF detector.
A logarithmic normal distribution was found to describe
the skewed deconvoluted spectra better than a symmetric Gaussian
and was used to find the position of the maximum of each spectrum.
Hence, each UCN production rate, which is in fact an integral UCN
production rate of the specific CN spectrum of this velocity
selector setting, was assigned to one wavelength $\lambda$.
The uncertainty of the wavelength measurement
was estimated as $\tfrac{\Delta \lambda}{\lambda} = \pm 1$~\%.

\section{Results}

\begin{figure}
    \centering
    \includegraphics[width=0.7\linewidth]{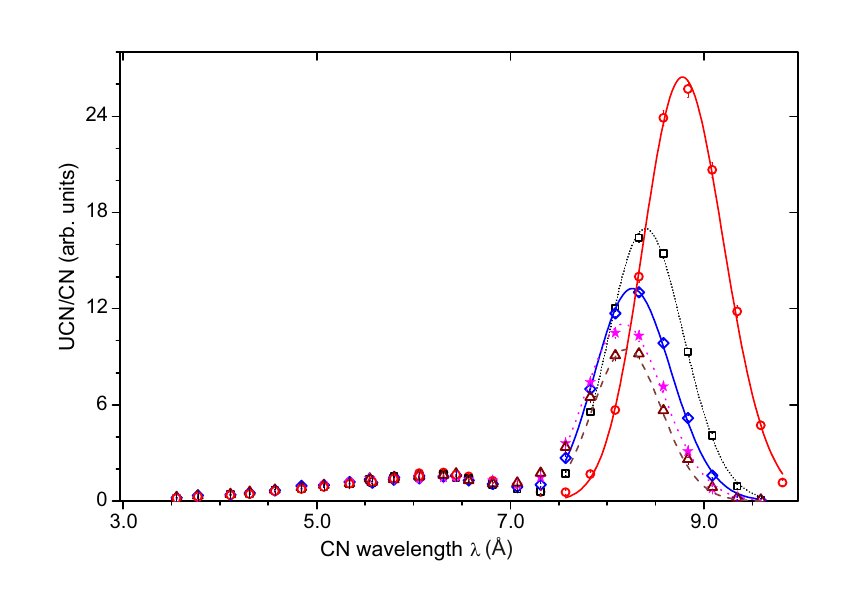}
    \caption{(Color online) Measured wavelength resolved UCN production rate per incident CN for
      $p=$\,SVP({\color{Red}$\ocircle$}), 6\,({\color{Black}$\square$}), 11\,({\color{Blue}$\Diamond$}),
    16\,({\color{Magenta}$\bigstar$}), 20~bar\,({\color{Brown}$\triangle$}). The lines are best fits of logarithmic normal distributions to the data with
    a reduced $\chi^2 = 1.3, 1.2, 2.0, 1.9, 2.6$, respectively.}
    \label{fig:peak_pressure}
\end{figure}

The measured wavelength-dependent UCN rates for all pressures,
normalized to the incident particle flux, are shown in
Fig.\,\ref{fig:peak_pressure}. 
Two different regions can be
distinguished for all pressures: a broad distribution for short
incident wavelengths from UCN production by multiphonon processes $R_{II}$
and a pronounced peak at long wavelengths from UCN production from
single-phonon excitation $R_I$ in He-II. 
These regions are separated by a minimum which gets less pronounced with pressure at the available wavelength resolution.
Note that the wavelength resolution was limited by the resolution of
the velocity selector (the same for the range of all single-phonon peaks shown).
With pressure the single-phonon peak
decreases in intensity and moves towards shorter wavelengths.

The area under the peak is proportional to the UCN
production rate by single-phonon excitation. 
The apparent peak width comes from the convolution of the incident CN spectra with
the extremely narrow single-phonon scattering function (the linewidth of the
roton excitation in $^4$He at $\sim$\unit[1]{K} is of the order of \unit[1]{\micro eV}, see e.g.\ Ref.\,\cite{Keller2004}, compared to the resolution of $\sim\unit[230]{\ueV}$ of the velocity selector at \unit[8]{\AA}).
The peak was fitted with a logarithmic normal distribution in $q$-space. This
ansatz fitted best the CN spectra and was used here again since
the single-phonon UCN production rate is proportional to the CN
intensity at $\lambda^{\ast}$ for each incident CN spectrum.
Note that fits with logarithmic normal distributions gave for all
pressures reduced $\chi^2$ values closer to unity than a symmetric Gaussian
distribution.
The fitting range was restricted to wavelengths larger
than $7.5$~{\AA}, hence excluding contributions from multiphonon
processes. 
The positions of the maximum of each fit function were observed for all pressures at
values smaller than expected from the crossing point of the
dispersion relations of He-II and the free neutron.  
They are listed in Tab.\,\ref{tab:productionRates} together with the UCN production
rates normalized to the rate at SVP\@. All single phonon UCN production rates $R_I$ were taken from the integrals over the fitted curves for $\lambda>\unit[7.5]{\text{\AA}}$. 

\begin{figure}
    \centering
		\subfloat{
    \includegraphics[width=0.47\linewidth]{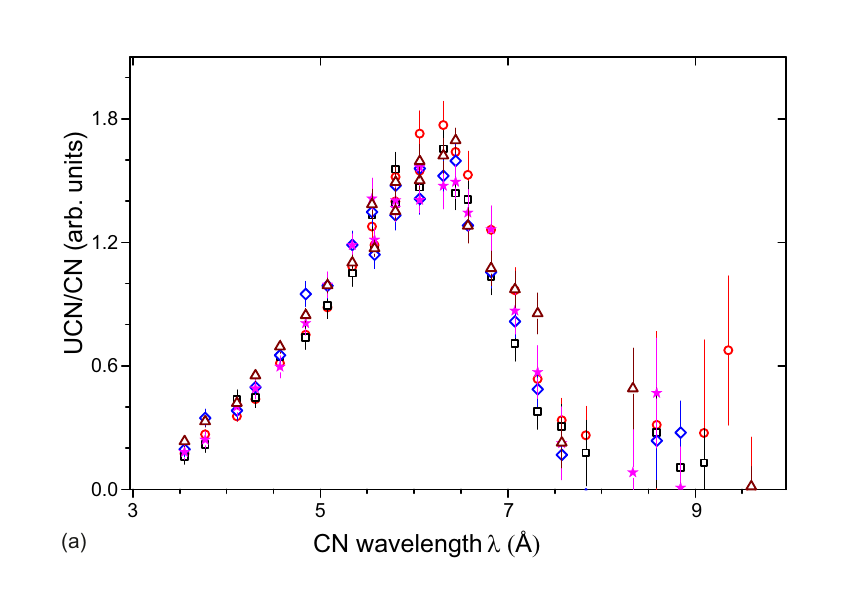}}
		\subfloat{
		\includegraphics[width=0.47\linewidth]{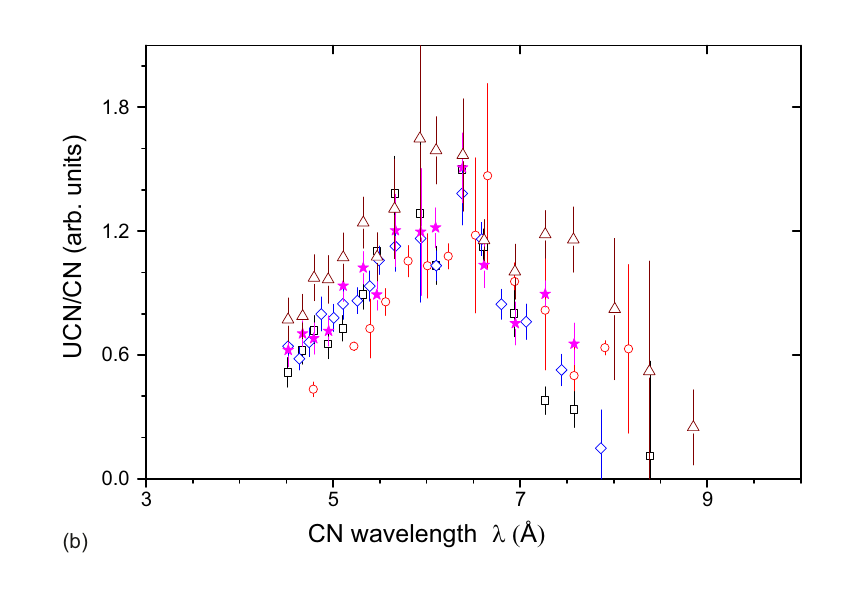}}

    \caption{(Color online) Ultracold neutron production per incident CN from multiphonon processes
      for $p=$\,SVP({\color{Red}$\ocircle$}), 6\,({\color{Black}$\square$}),
      11\,({\color{Blue}$\Diamond$}), 16\,({\color{Magenta}$\bigstar$}),
      20~bar\,({\color{Brown}$\triangle$}). Measured data~(a) after subtracting the
      single-phonon part (see text). Calculated~(b) from inelastic scattering data for $p=$\,SVP({\color{Red}$\ocircle$}), 5\,({\color{Black}$\square$}),
      10\,({\color{Blue}$\Diamond$}), 15\,({\color{Magenta}$\bigstar$}), and
      20~bar\,({\color{Brown}$\triangle$}) .}
    \label{fig:ExpTheory}
\end{figure}

The obtained fits were then
subtracted from the data in order to extract the UCN production
rates due to multiphonon processes.
These are shown scaled up in Fig.\,\ref{fig:ExpTheory} and found to exhibit a broad, pressure-independent maximum at approximately \unit[6.25]{\text{\AA}}.
For estimating the relative change of UCN production due to multiphonon excitation and multiple scattering with pressure we used a simple sum over data points for $ \lambda\!<\!\unit[7.5]{\text{\AA}}$, which was
found to remain constant within errors in this range (see
Table\,\ref{tab:productionRates}).

\tsdel{
Note that this approach provides a direct relative comparison of the
production rates at different pressures: For a fixed wavelength, the
same incident beam was used. Pressure-dependent UCN losses (e.g.\ due
to cracks in the coating of the storage vessel) would affect the same way the measurement of single-phonon production. 
}

\section{Discussion}

\begin{figure}
    \centering
    \includegraphics[width=0.7\linewidth]{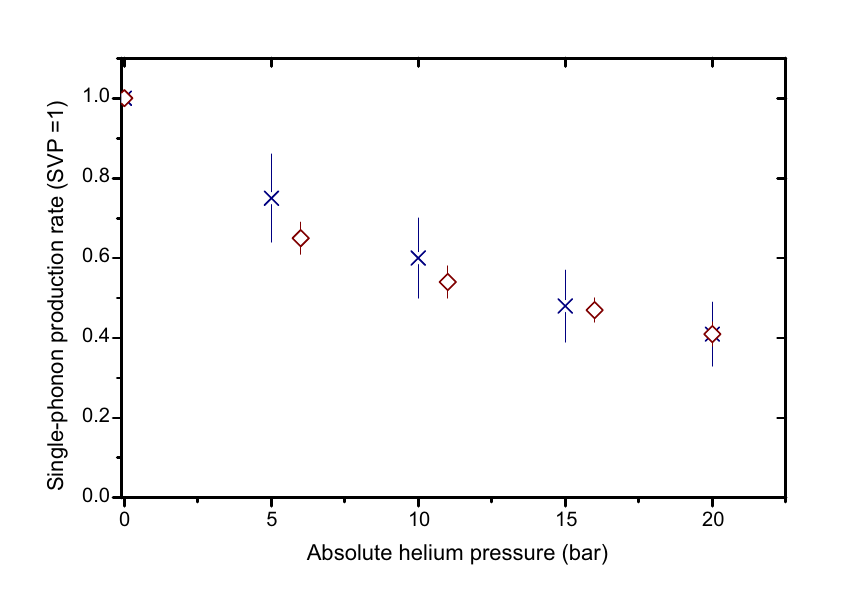}
    \caption{(Color online) Comparison of the pressure dependence of single-phonon UCN production from
      this experiment ({\color{Brown}$\Diamond$}) and from calculation from scattering
      data ({\color{Blue}$\times$})\,\cite{Schmidt-Wellenburg2009,Schmidt-WellenburgPhD}. Both dependencies have
      been normalized to the respective production rate at SVP\@.}
    \label{fig:PressureDependenceProductionRates}
\end{figure}

In our experiment we observed the anticipated decrease of UCN production $R_I$ due to single-phonon processes with pressure. This behavior can be well understood by taking a look at

\begin{equation}
     R_I \propto N \beta \lambda^{\ast 4} S^{\ast}\left. \frac{\mrm{d}\phi}{\mrm{d}\lambda} \right\vert_{\lambda^{\ast}},
\label{P_I}
\end{equation}

\noindent as derived in an unpublished note of Pendlebury (see also \cite{Golub1991}), where the single-phonon scattering intensity $S^{\ast}=S(q^{\ast},\omega^{\ast}$) decreases from $S^{\ast}(\mrm{SVP}) = 0.118(8)$ to
$S^{\ast}(20~\mrm{bar}) = 0.066(6)$ \cite{Schmidt-Wellenburg2009}.
This change of the structure factor explains most of the observed decrease. 
Smaller contributions come from the change of $\lambda^{\ast}$ and from reduced overlap
between the dispersion relations of He-II and of the free neutron
taken into account by the parameter $\beta(\text{SVP}) = 1.42$ to $\beta(\unit[20]{bar}) = 1.21$ \ts{(values deduced from the scattering data \cite{Gibbs1999}, see \cite{Schmidt-WellenburgPhD})}. The increase in helium
number density $N [10^{22}\mrm{cm^{-3}}]$ from $\tsdel{N_{\text{SVP}}}\ts{N(\text{SVP})}
=2.183\tsdel{5}\ts{6}$ to $\tsdel{N_{\text{20\,bar}}}\ts{N(\unit[20]{bar})} =2.53\tsdel{17}\ts{49}$
\ts{(calculated from the mass densities at \unit[0.1]{K} from \cite{Abraham1970})} cannot
compensate these effects.
Note that the factor $\left.\frac{\mrm{d}\phi}{\mrm{d}\lambda} \right\vert_{\lambda^{\ast}}$
does not contribute to the normalized data in Fig.\,\ref{fig:peak_pressure}.
Its inclusion for a typical CN spectrum does also not compensate the
discussed decreases. 
This behavior confirms the results from calculations, which agree within errors with the experimental results, see Tab.\,\ref{tab:productionRates} and Fig.\,\ref{fig:PressureDependenceProductionRates}. 

We observe a displacement of all measured single-phonon production rate peaks to lower wavelengths than anticipated from the crossing points of the free neutron dipsersion curve with the one of He-II\@. This can be explained by an unintentional displacement of TOF monitor by \unit[5]{cm}, the width of the detector. This most probably has happened during the setup for the second measurement period (the only one which delivered data). For this reason we present all our results in table\,\ref{tab:productionRates} using a ``corrected'' wavelength scale $\lambda_\text{cor} =1.025\!\cdot\!\lambda$.

\begin{table}
%
				%
		%
		%
%

\begin{tabular}{|r|r@{.}lr@{.}lr@{.}lr@{.}l||r|r@{.}lr@{.}lr@{.}l|}\hline
        & \multicolumn{8}{c||}{experiment } & & \multicolumn{6}{c|}{from Ref.\,\cite{Schmidt-Wellenburg2009,Schmidt-WellenburgPhD}}  \\
				
        \multicolumn{1}{|c|}{$p$}  & \multicolumn{2}{c}{$\lambda^{\ast}$} & \multicolumn{2}{c}{$\lambda^{\ast}_{\text{cor}}$} & 
	  \multicolumn{2}{c}{s-ph}  &\multicolumn{2}{c||}{m-ph} & \multicolumn{1}{|c|}{$p$} & \multicolumn{2}{c}{$\lambda^{\ast}$} &\multicolumn{2}{c}{s-ph}  &
	  \multicolumn{2}{c|}{m-ph} \\
		
        \,[bar]  & \multicolumn{2}{c}{[\AA]} & \multicolumn{2}{c}{[\AA]} & \multicolumn{2}{c}{} & \multicolumn{2}{c||}{$\lambda\!<\!\unit[7.5]{\text{\AA}}$} &\,[bar]&
	  \multicolumn{2}{c}{[\AA]} & \multicolumn{2}{c}{} & \multicolumn{2}{c|}{$\lambda\!<\!\unit[7.5]{\text{\AA}}$} \\ \hline
		
        SVP & 8&70(1) & 8&92(set) &  \multicolumn{2}{l}{1} & \multicolumn{2}{l||}{1} & SVP & 8&92(2)& \multicolumn{2}{l}{1}  & \multicolumn{2}{l|}{1}\\ \hline
        6   & 8&31(1) & 8&52(1)    & 0&65(4)& 0&93(4)&  5  & 8&54(4)& 0&75(11) & 1&09(3)\\ \hline
        11  & 8&18(1) & 8&39(1)    & 0&54(4)& 0&97(4)& 10  & 8&39(2)& 0&59(10)  & 1&12(3)\\ \hline
        16  & 8&10(1) & 8&30(1)    & 0&47(3)& 0&97(4)& 15  & 8&30(2)& 0&47(9) & 1&17(6)\\ \hline
        20  & 8&09(1) & 8&29(1)    & 0&41(3)& 1&03(4)& 20  & 8&26(2)& 0&40(8) & 1&43(6)\\ \hline
        \end{tabular}
        \caption{Pressure dependence of measured single-phonon peak position
	$\lambda^{\ast}$, corrected values $\lambda^{\ast}_{\text{cor}}$ (see text), and UCN
	production rates compared with results given in Ref.\,\cite{Schmidt-Wellenburg2009,Schmidt-WellenburgPhD}
	calculated from scattering data.
	The UCN production rates are normalized to the incident cold neutron flux.
  The single-phonon~(s-ph) and multiphonon~(m-ph) columns show the production rates
	at pressure $p$ normalized to the respective SVP
	results.
	The quoted errors of $\lambda^{\ast}$ from this experiment only include the uncertainties
	from the fits with the logarithmic normal distribution.
	The quoted errors of the measured UCN production rates are statistical uncertainties; in the single-phonon
	case uncertainties of the fit were increased to yield a reduced $\chi^2$ of 1.
        The errors of the calculated values are the quadratic sums of all contributions.}
        \label{tab:productionRates}
    \end{table}

In the multiphonon range we observed a discrepancy between the measured and the calculated \ts{pressure dependence of the} UCN production rate (see Tab.\,\ref{tab:productionRates} and Fig.\,\ref{fig:ExpTheory}).
From calculations we would have expected an increase with pressure, whereas no increase could be observed\tsdel{ in the relevant range}.
The observed pressure independence of the storage time constant does not indicate a change of the UCN extraction probability from the converter vessel with pressure.
Note that our experimental approach provides a direct relative comparison of the
production rates at different pressures: For a fixed wavelength, the
same incident beam was used.
Pressure-dependent UCN losses (e.g.\ due
to cracks in the coating of the storage vessel) are independent of the CN wavelength.

One explanation for the discrepancy in the multiphonon region
could be a change in the dynamic
structure factor $S(q,E)$ with temperature. The calculations
Ref.\,\cite{Schmidt-Wellenburg2009,Schmidt-WellenburgPhD} employ scattering
data measured at \unit[0.5]{K} whereas our measurements were done at \unit[1.1]{K}. Increasing
temperature results, for example, in a broadening of the single-phonon
excitation. This increase is small at low temperatures but becomes large
towards the $\lambda$-transition\,\cite{Gibbs1999}. In\,\cite{Gibbs1999} the
temperature dependence of the single-phonon excitation linewidth for different pressures was determined from fits to the measured dynamic
structure factor. 
No significant change of these values was found below about \unit[1.3]{K}. Likewise, the multiphonon 
scattering varies only weakly with temperature\,\cite{Gibbs1999}. Therefore we consider the
scattering data measured at \unit[0.5]{K} as adequate to predict UCN production at \unit[1.1]{K}.
Note that temperature also influences UCN up-scattering but this effect
cannot explain the observed discrepancy as it is independent on the CN
wavelength, and would also shift the single-phonon data from expectation.

Assuming no significant change of the dynamic structure factor between \unit[0.5]{K} and
\unit[1.1]{K}, the observed discrepancy can only be explained by an
effect depending on both, the CN wavelength and the He pressure, such as
inelastic scattering to other than UCN energy. Inelastic scattering
has two types of consequences: (i)~reduction of the beam intensity averaged
over the production volume with respect to the incident one, as scattered
neutrons may leave the production volume sideways; and (ii)~change in CN
energy which may influence the probability of UCN production for the
scattered neutron (multiple scattering). Both effects are not taken into
account in the calculations.


In order to estimate the reduction of the beam intensity (i) we use
the wavelength- and temperature-dependent transmission of neutrons through
liquid He at SVP measured by Sommers~et~al.\,\cite{Sommers1955}.
We calculate the volume-average CN~flux in the
multiphonon range (from 3.5~\AA{} to 7.5~\AA{}) inside the production
volume relative to the incoming flux:

\begin{equation}
t_{\text{m-ph}}=\int_{\text{m-ph}}
\frac{\mrm{d}\phi}{\mrm{d}\lambda}
t(\lambda)\,\mrm{d}\lambda\left/\int_{\text{m-ph}}
\frac{\mrm{d}\phi}{\mrm{d}\lambda}\, \mrm{d}\lambda\right. .
\label{eq:6}
\end{equation}

\noindent Values between 3.5~{\AA} and 4.52~{\AA}
were obtained from linear extrapolation of the data given there.
We obtain $t_{\text{m-ph}}\ts{(\text{SVP})}=0.89$.
For 8.9~{\AA}, the correction is $t_{\text{s-ph}}\ts{(\text{SVP})}=0.98$. Note that
for a longer production volume, the correction would be larger.
We define multiphonon and single-phonon production rates, $R_{II}$
and $R_{I}$, as integrals over the respective regions in the
CN-normalized UCN production spectrum shown in Fig.\,\ref{fig:peak_pressure}. In order to compare the observed ratio $R_{II}/R_I\,\big(\text{SVP}\big)=0.14(1)$
with the calculations,
we need to correct the prediction with the geometry specific attenuation:
$t_{\text{m-ph}}/t_{\text{s-ph}}\cdot
R^{\text{calc}}_{II}/R^{\text{calc}}_I\,\big(\text{SVP}\big)=0.11(1)$.

A reduction of CN transmission $t_{\text{m-ph}}$ with increasing pressure
could explain the observed discrepancy between the measured and the
calculated pressure dependence of multiphonon UCN production. Unfortunately,
a quantitative confirmation of this hypothesis is not possible, due to
lacking transmission data for higher pressures. Also scattering data
for different pressures \cite{Gibbs1999} do not cover the high-$q$ range
needed to calculate the transmission via total scattering.

The change of the CN energy by inelastic scattering (ii) depends on the
incident wavelength and the pressure. Scattering to the wavelength $\lambda^{\ast}$
is kinematically possible and would increase the probability of UCN production by the
scattered CN; however, the width of single-phonon UCN production
is very small. Multiple scattering was not taken into account in the calculations
\cite{Schmidt-Wellenburg2009,Schmidt-WellenburgPhD}. Its contribution
to the UCN production in our setup was estimated by Monte Carlo simulations\,\cite{Farhi2013}
based on the McStas package\,\cite{Lefman1999,Willendrup2004}.
For SVP, multiple scattered CN were found to contribute about \unit[6]{\%} to the UCN production rate
in the multiphonon region. This fraction decreases slightly with increasing pressure, to \unit[5]{\%} for
$p\ge\unit[10]{bar}$. The weak pressure dependence of this small contribution
cannot explain the discrepancy between observed and expected pressure dependence of multiphonon UCN production.
It changes the calculated relative production rates at SVP $t_{\text{m-ph}}/t_{\text{s-ph}}\cdot
R^{\text{calc}}_{II}/R^{\text{calc}}_I\,\big(\text{SVP}\big)$ to \unit[0.12(1)].

However, it should be noted that the accuracy of the simulations and calculations
is limited by the available scattering data: Single data sets cover only a limited
$(q,\omega)$-range or provide a limited resolution. Therefore, data sets
\cite{Bossy2009,Bossy2000,Gibbs1995} from different instruments and with different
resolution settings had to be combined in the simulations \cite{Farhi2013}. The scattering
data \cite{Gibbs1999} used for the calculations
\cite{Schmidt-Wellenburg2009,Schmidt-WellenburgPhD} were obtained
with two different incident wavelengths at IN6: \unit[4.1]{\AA} for SVP and \unit[4.6]{\AA}
for higher pressures. 
Possibly uncorrected effects may contribute to the observed discrepancies.

\section{Conclusion}
We have investigated CN-wavelength-dependent UCN production in
pressurized He-II for pressures up to $20$~bar. The single-phonon
UCN production peak moves to shorter wavelength and decreases in
intensity, whereas the absolute contribution from multiphonon
processes stays constant within the experimental uncertainties.
We have found that predictions from
calculations based on inelastic neutron scattering
data\,\cite{Schmidt-Wellenburg2009,Schmidt-WellenburgPhD} agree well
with the behavior of the single-phonon UCN production rate. However,
the same calculations disagree with our measurement in the
multiphonon range. There is no indication for a significant change
in the dynamic structure factor between \unit[0.5]{K} and \unit[1.1]{K} that could
invalidate the calculations for our experiment. However, the disagreement
could be caused by a pressure- and wavelength-dependent attenuation of the CN beam in
the He-II.
This hypothesis could not be tested due to a lack of
transmission data at higher He pressure. Note that the ratio of
multiphonon to single-phonon UCN production would reduce for longer
production volumes. The calculations also did not take into account
modifications of the CN spectrum by inelastic scattering inside the
He-II, but Monte-Carlo simulations showed that the contribution
of UCN production by multiple scattered CN and its pressure dependence
are too small to explain the observed disagreement.

The disagreement of the position of the single-phonon peak at SVP
with earlier measurements by Yoshiki and
coworkers\,\cite{Yoshiki1992} and Baker and
coworkers\,\cite{Baker2003} can most probably be explained by an
accidental displacement of the TOF detector in our experiment.

In general, the relative UCN production rate due to single-phonon and
multiphonon processes depends upon the incident CN
spectrum. For the spectrum in our experiment we have found that
\unit[$(30\pm2)$]{\%} of the total UCN production over all
incident wavelengths are from multiphonon processes. Baker and coworkers\,\cite{Baker2003}
found a multiphonon contribution of $(24\pm2)\%$
for a different CN beam.
Note that, according to our previous
discussion, also the length of the UCN production volume
(which was \unit[326]{mm} in the experiment of Baker and coworkers)
influences this ratio.

Using the entire spectrum of a white beam might still be advantageous for a powerful UCN source, as
monochromatization by a crystal not only cuts out a narrow wavelength band around
$8.92$~\AA{}, but also reduces the intensity of the Bragg reflected
beam substantially
(factor 2-5 \cite{Schmidt-Wellenburg2009,Piegsa2014}).
In whatever
case, monochromatic or white incident beam, best results are
obtained for He-II at SVP; without even taking into account the
increase in technical complexity when going to higher pressures. 
For future searches of the neutron electric dipole moment it might still
be beneficial to apply a pressure of a few \unit[100]{mbar} to
increase the dielectric strength without
loosing significantly in UCN production.
A future perspective to increase UCN production might be the use of
solid helium as UCN converter, which was not possible to investigate
during this experimental run.

We would like to thank the ILL staff, especially D.~Berruyer, T.~Brenner,
P.~Lachaume, P.~Mutti, and P.~Thomas who made this
experiment possible thanks to their technical support. We would like
to thank S.~Mironov for his important contributions to the design of
the cryogenic apparatus. This work has been funded by the German
BMBF~(contract number 06MT250).


\end{document}